\documentclass[namedreferences]{SolarPhysics}
\usepackage[optionalrh]{spr-sola-addons} 
\usepackage{graphicx}                    
\usepackage{courier}                    
\usepackage{amssymb}                    
\usepackage{color}                       
\usepackage{url}                         

\newcommand{\pref}{\protect\ref}
\newcommand{\trace}{{\em TRACE}}
\newcommand{\soho}{{\em SOHO}}
\newcommand{\hinode}{{\em Hinode}}

\newcommand\arcsec{\mbox{$^{\prime\prime}$}}

\begin{document}

\begin{article}

\begin{opening}

\title{The Impact of New EUV Diagnostics on CME-Related Kinematics}

%
\author{Scott W.~\surname{McIntosh}$^{1}$\sep Bart~\surname{De Pontieu}$^{2}$\sep Robert J.~\surname{Leamon}$^{3}$}

%
\runningauthor{McIntosh, De Pontieu \& Leamon}
\runningtitle{New EUV Diagnostics and CMEs}

%
\institute{$^{1}$ High Altitude Observatory, \\National Center for Atmospheric Research,\\  P.O. Box 3000, Boulder, CO 80307, USA.\\ email: \url{mscott@ucar.edu}\\ 
                $^{2}$ Lockheed Martin Solar and Astrophysics Lab,\\ 3251 Hanover St., Org. ADBS, Bldg. 252,\\ Palo Alto, CA  94304, USA.\\ email: \url{bdp@lmsal.com}\\
                $^{3}$	 Department of Physics, Montana State University, Bozeman, MT 59717, USA.\\ email: \url{robert.j.leamon@nasa.gov}\\}

\date{Received ; accepted}

\begin{abstract}
We present the application of novel diagnostics to the spectroscopic observation of a Coronal Mass Ejection (CME) on disk by the Extreme Ultraviolet Imaging Spectrometer (EIS) on the \hinode\/ spacecraft. We apply a recently developed line profile asymmetry analysis to the spectroscopic observation of NOAA AR~10930 on 14-15 December 2006 to three raster observations before and during the eruption of a 1000~km/s CME. We see the impact that the observer's line-of-sight and magnetic field geometry have on the diagnostics used. Further, and more importantly, we identify the on-disk signature of a high-speed outflow behind the CME in the dimming region arising as a result of the eruption. Supported by recent coronal observations of the {\em STEREO} spacecraft, we speculate about the momentum flux resulting from this outflow as a secondary momentum source to the CME. The results presented highlight the importance of spectroscopic measurements in relation to CME kinematics, and the need for full-disk synoptic spectroscopic observations of the coronal and chromospheric plasmas to capture the signature of such explosive energy release as a way of providing better constraints of CME propagation times to L1, or any other point of interest in the heliosphere. 
\end{abstract}

%

\end{opening}

%
\section{Introduction}

In this short forward-looking and speculative paper, we present an extended analysis of observations of NOAA AR~10930 from the Extreme-ultraviolet Imaging Spectrometer (EIS; \opencite{2007SoPh..243...19C}) on {\em Hinode} \cite{2007SoPh..243....3K} between 19:00~UT December 14 2006 and 06:00~UT December 15 2006. This time period saw an X-Class flare and a $\sim$1000~km/s halo CME\footnote{The CME properties were automatically derived from \soho{}/LASCO data by the NASA/GSFC CDAW (\url{http://cdaw.gsfc.nasa.gov/}) and the Royal Observatory of Belgium/SIDC CACTUS (\url{http://www.sidc.be/cactus/}; \opencite{2004AA...425.1097R}) catalogues.} and coronal dimming event (e.g., \opencite{2001JGR...10629239K}, \opencite{2008SoPh..252..349A}) that emanated from this complex active region at around 20:12~UT.

We expand on the analysis of \inlinecite{2007ApJ...660.1653M}, \inlinecite{2009ApJ...693.1306M}, and \inlinecite{2009arXiv0901.2817M}, exploiting the rare detailed spectroscopic measurements of dimming region evolution that were first studied by \inlinecite{2007PASJ...59S.801H}. EIS provides a tantalizing look at the dynamic behavior of EUV emission lines over the course of the eruption. The interpretation of the dynamic evolution of the non-thermal line widths presented forms a extension of the challenge posed by \inlinecite{2009ApJ...693.1306M} to the rapidly increasing sophistication of numerical CME models: specifically that they {\em need\/} to cope with the complex thermodynamics of the CME source region. Clearly, the relationship between the dimming region and CME is one that grows considerably when only narrowband spectroscopic observations are considered. Therefore, rigorously establishing the poorly understood physical connection between CMEs and coronal dimmings using detailed spectroscopic measurement is a must. Fortunately, there is an ever-growing list of investigations in the literature (\opencite{2000A&A...358.1097H}; \opencite{2007PASJ...59S.793I}; \opencite{2007PASJ...59S.801H}; \opencite{2008A&A...478..897B}; \opencite{2009ApJ...702...27J}; \opencite{2009AdSpR..44..446H}) that, we hope, will continue to expand with ongoing EIS observation and the launch of IRIS, the Interface Region Imaging Spectrograph\footnote{See the IRIS website (\url{http://iris.lmsal.com/}) for more information about the mission.}, in late 2012.

In the following sections we describe the observations used and the technique developed to assess the asymmetry of the emission line profiles 
(e.g., \opencite{2009ApJ...701L...1D}; \opencite{2009arXiv0910.2452M}), new imaging results that validate the spectral analysis, how we interpret those datasets, and finally, speculate on the implications of the results on the kinematic properties of the CME itself. 

%
%
\begin{figure} 
\centerline{\includegraphics[width=1.0\textwidth,clip=true]{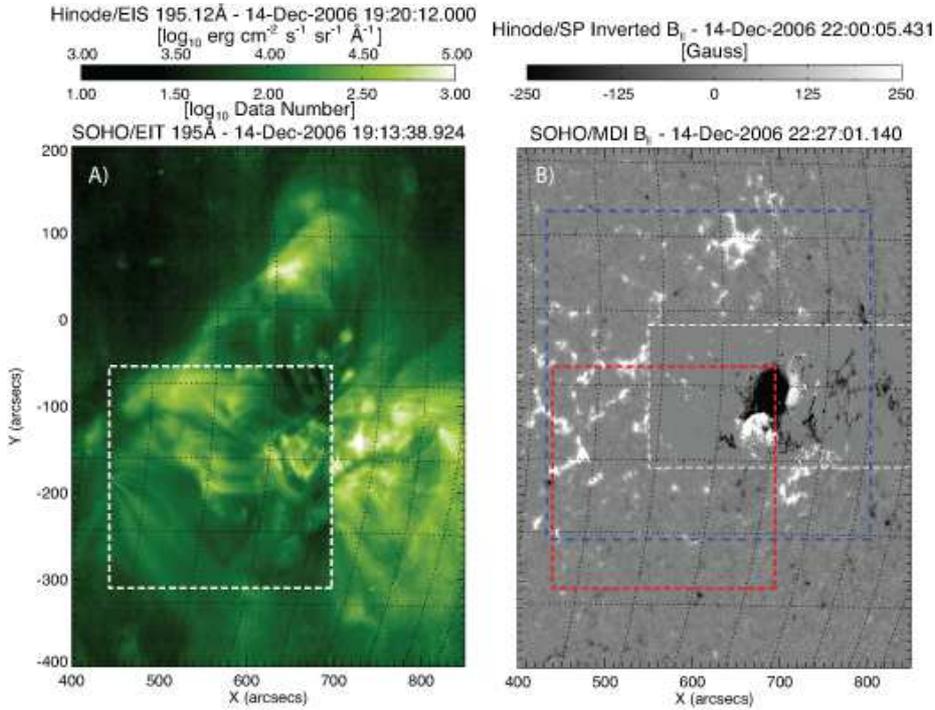}}
\caption{Representative pre-CME images of NOAA AR 10930 from \soho\ and \hinode{}. From left to right we show the full-field EIT 195\AA{} and MDI LOS magnetogram with \hinode{}/EIS and \hinode\ SOT/SP images inlaid for reference respectively. The red and blue dashed regions in the right panel respectively show the EIS field of view and region of the composite MDI/SP magnetogram extrapolation used to provide context below.}
\label{fig1}
\end{figure}

\section{Observations \& Analysis}\label{s:method}
The dataset of interest comprises three spectroheliogram ``raster'' observations (19:20-21:34~UT, 01:15-03:30, 04:10-06:24~UT), targeted at the trailing
edge of the active region that is the source of the event studied. The EIS observations are reduced using the IDL Solarsoft (\opencite{1998SoPh..182..497F}) {\tt eis\_prep} algorithm which corrects for cosmic ray hits, hot pixels, detector bias, and dark current, and converts data numbers to intensities (in erg~cm$^{-2}$~s$^{-1}$~sr$^{-1}$~\AA$^{-1}$). 

Each EIS raster is comprised of 256 horizontal (West to East) mirror mechanism steps with the $1\arcsec$ slit at a spacing of $1\arcsec$ and a height of $256\arcsec$ and has information in nine 24-pixel wide spectral windows. At a spectral resolution of 22.3m\AA{}, and wavelength of 195\AA{}, one pixel on the detector is equivalent to a velocity of $\sim$34~km/s. Panel~A of Fig.~\pref{fig1} shows the region surrounding AR~10930 provided by the 195\AA{} passband image of {\soho\/}/EIT (\opencite{1995SoPh..162..291D}) taken at 19:13~UT which is inlaid with the peak line intensity of the EIS 195\AA{} raster.

The SOT Spectro-Polarimeter (SP) rastered the region of interest twice during the time of interest (14 December 2006 22:00-23:03~UT and 15 December 2006 05:45-06:48) taking 1000 stepped measurements per raster in Stokes I, Q, U \& V in Fe~I 6301.5, 6302.5\AA{} and binning two-by-two pixels on board to give an effective (linear) spatial scale of $0.32\arcsec$. The SP data were reduced using the standard settings of the {\tt sp\_prep} routine in the SOT software tree. The Stokes polarization signals measured by SP are inverted into a set of physical parameters that describe the vector magnetic field (field strength $B$, inclination from the local normal $\psi$, azimuth $\phi$, filling factor $S_{\alpha}$, etc) and local plasma conditions using the approach of \inlinecite{1987ApJ...322..473S}: seeking to simultaneously minimize the least-squares fit of four Stokes profiles with analytic descriptions of the polarization signals under the Zeeman effect in a Milne-Eddington atmosphere. The Stokes inversion shown here is executed using the MERLIN algorithm (\opencite{2007MmSAI..78..148L}) that has recently been developed at NCAR under the framework of the Community Spectro-polarimetric Analysis Center (CSAC; \url{http://www.hao.ucar.edu/projects/csac/}) for use with SP. The full SP Stokes vector maps are shown, for reference, in \inlinecite{2009arXiv0901.2817M}.

An example of an SP line-of-sight (LOS) magnetogram, constructed from the inverted Stokes I, Q, U \& V measurements, allows us to look in detail at the LOS field: $B_{||} = S_{\alpha} B\cos\psi$ and is inlaid in the full-disk LOS magnetogram from \soho{}/MDI (\opencite{1995SoPh..162..129S}) and shown in panel B of Fig.~\pref{fig1}. Note that we have used the full-disk \soho\ images/magnetograms to align the sub-field \hinode\ observations of EIS and SOT---this allows us to correct the pointing of the \hinode\ data to within $3\arcsec$.  Indeed, using the \soho\ data as a pointing reference has the added advantage of allowing good coalignment between EIS and SOT. In Fig.~\pref{fig1}B, the subfield covered by EIS is indicated by a red dashed outline while the blue dashed outline marks a region for which we extract a potential field extrapolation for the purpose of later discussion (see, Fig.~\ref{fig4}).

\subsection{R-B Analysis}

Following the description of \inlinecite{2009ApJ...701L...1D} and \inlinecite{2009arXiv0910.2452M} we perform a `Red Minus Blue' (R-B) profile asymmetry analysis on spectral lines in the EIS data that are not significantly impacted by spectral blends in the relatively narrow (24 pixel) spectral windows used, i.e., the Fe~XIII 202\AA{} and Fe~XIV 274\AA{} lines. The R-B analysis involves several steps. First we fit a single Gaussian shape to the emission line profile at each pixel to establish the line centroid. Once determined, we sum the amount of emission in narrow ($\sim$24~km/s wide) spectral regions symmetrically placed about the determined centroid in a line profile interpolated to ten times the spectral resolution. We then subtract the red and blue wing contributions to the interpolated profile to make a filtergram sampling a particular velocity range. A positive value of R-B indicates an asymmetry in the red wing of the line, which we can interpret as the signature of excess downflowing material at that velocity while, conversely, a negative value of R-B would indicate an excess of upflowing material.

\begin{figure} 
\centerline{\includegraphics[width=1.0\textwidth,clip=true]{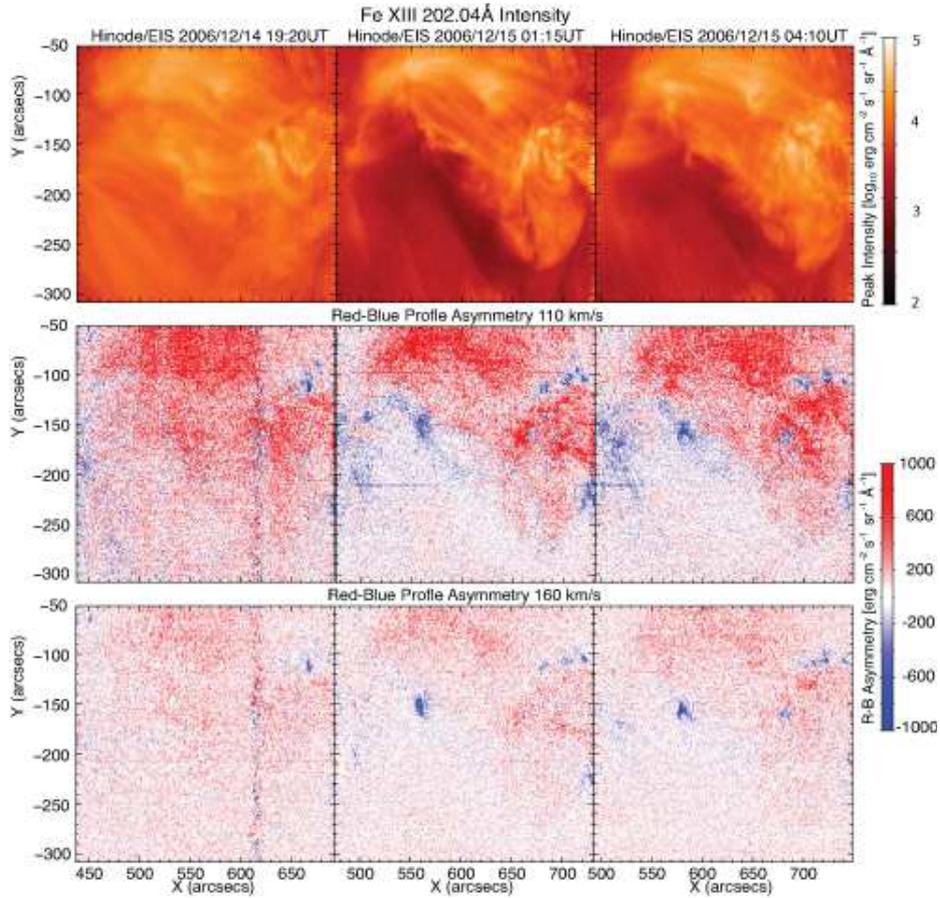}}
\caption{Three phases of evolution of the CME as observed in the Fe~XIII 202\AA{} emission line peak intensity (top row) and R-B analysis maps (see text for details of the R-B method) showing the R-B maps at 100~km/s (middle row) and 160~km/s (bottom row). The color of the R-B  maps indicate the net direction of the plasma motion at that velocity, red regions suggest net downflow while blue regions suggest net upflow.}\label{fig2}
\end{figure}

\begin{figure} 
\centerline{\includegraphics[width=1.0\textwidth,clip=true]{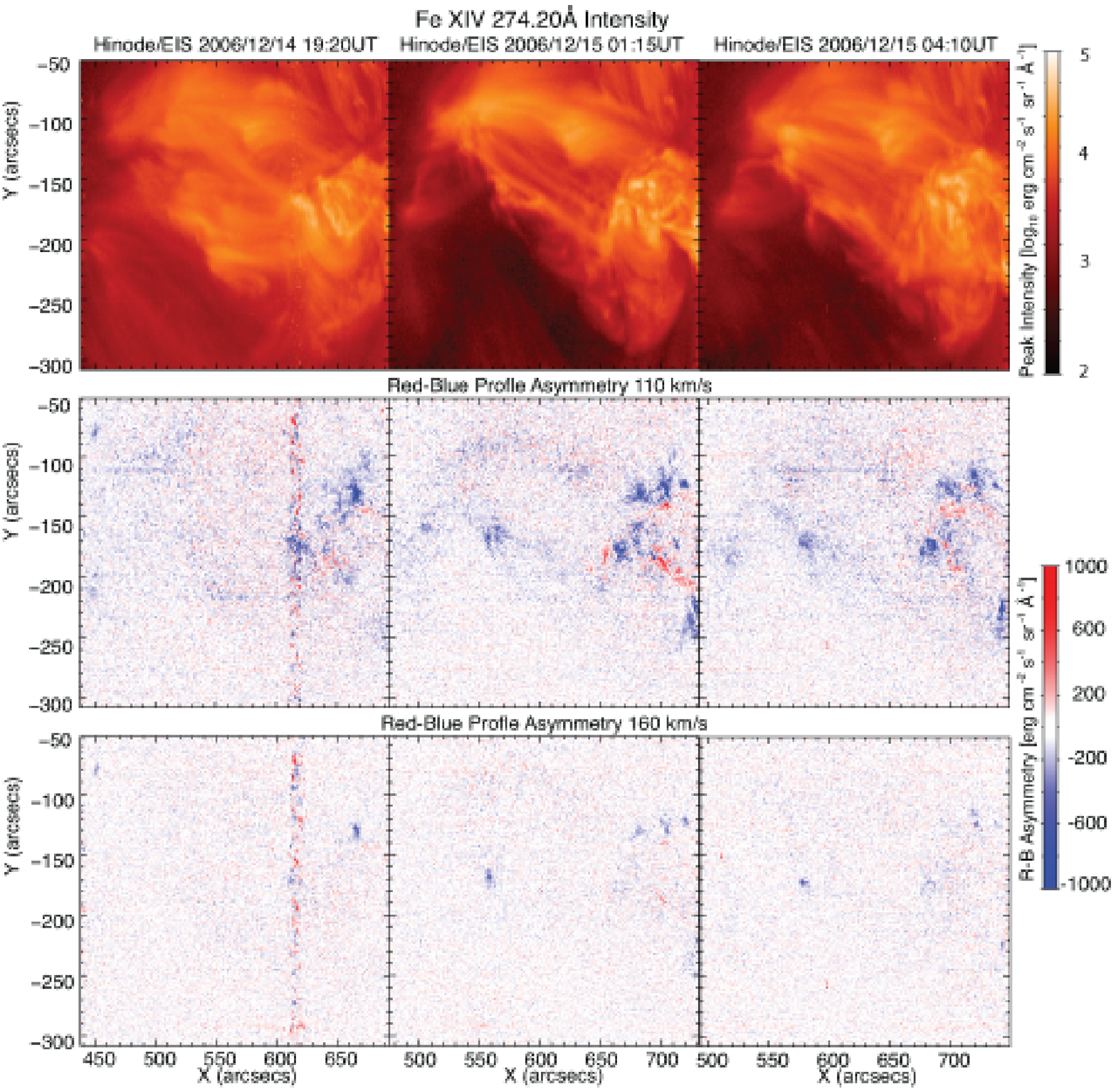}}
\caption{Three phases of evolution of the CME as observed in the Fe~XIV 274\AA{} emission line peak intensity (top row) and R-B analysis maps (see text for details of the R-B method) showing the R-B maps at 100~km/s (middle row) and 160~km/s (bottom row). The color of the R-B  maps indicate the net direction of the plasma motion at that velocity, red regions suggest net downflow while blue regions suggest net upflow. Movies 1, 2, and 3 of the online material are provided to support the three columns of the figure from left to right respectively.}\label{fig3}
\end{figure}

Figures~\pref{fig2} and~\pref{fig3} show the peak line intensities (top row), R-B at 110~km/s (middle row), and at 160~km/s (bottom row) for the Fe~XIII 202\AA{} and Fe~XIV 274\AA{} lines respectively for the three phases of the dimming observed (left column \-- 19:20-21:34~UT; middle \-- 01:15-03:30~UT; right \-- 04:10-06:24~UT). We note that movies of the complete velocity range are available online and the presence of the vertical anomaly in the pre-eruption spectroheliograms that were caused by a couple of bad detector read outs at a few slit positions.

The three movies supporting Fig.~\pref{fig3} (Movies 1 through 3) show the progression of the R-B analysis for the Fe~XIV 274\AA{} line from 40~km/s through 200~km/s in context with the peak line intensity, $1/e$ line width, and (relatively calibrated) Doppler velocity over the course of the dimming event observed by EIS. In all of the movies we notice the general correspondence between the regions of enhanced line broadening, significant blue wing asymmetries and the darkest coronal loop structures as was noted, and discussed, in \inlinecite{2009arXiv0910.2452M}. During the dimming event we notice the extension of the excess non-thermal broadening of the line into the regions far from the magnetic footpoints (and associated blue wing asymmetry) of the dimming region. It is this extended, and dynamic, variation of the line widths that \inlinecite{2009ApJ...693.1306M} attributed to the growth of Alfv\'en waves on the now open, less dense, coronal magnetic field lines behind the CME. However, \inlinecite{2009ApJ...701L...1D} and \inlinecite{2009arXiv0910.2452M} suggested that the blueward asymmetries in a wide variety of lines formed at transition region and coronal temperatures may be caused by upflows from chromospheric spicules that are associated with hot upflows at $\sim$100~km/s. Given this interpretation and our observation of excess blueward asymmetries where the linewidths are enhanced, we would expect that a significant portion of the excess broadening at the magnetic footprint of these regions is caused by the presence of high velocity spicular outflows. Correctly aportioning the observed broadening there to spicular outflows or larger amplitude Alfv\'enic motions of the roots of the magnetic field lines is difficult without higher temporal, spatial, and spectral resolution data. This is an issue we will return to below.

\section{Results}\label{s:results}
The temporal evolution of the line intensity are as described in \inlinecite{2007ApJ...660.1653M} and \inlinecite{2009ApJ...693.1306M}, where in the central panels of the top row we see a sudden dimming to the South-East of the active region between the first and second rasters; there is a $\sim$75\% reduction in intensity in that region between the first two images. By the third raster, we see some filling of the region has taken place close to the active region \cite{2007ApJ...660.1653M}. We see that the bulk of the dimming originates in the loop complexes that originate in the positive polarity flux domains at $[x, y]=[450\arcsec, -170\arcsec]$ (location 1) and $[530\arcsec, -150\arcsec]$ (location 2) in Fig.~\pref{fig1}. These loop systems span the South-Eastern portion of the active region connecting it to the negative polarity flux at $[650\arcsec, -160\arcsec]$, the sunspot, and surrounding flux. Aspects of this connectivity are shown in Fig.~\pref{fig4}.

Prior to the eruption (the first columns of Fig.~\pref{fig2} and~\pref{fig3}) we see a very weak blue wing asymmetry in Fe~XIII at 110~km/s in location 1 that is fainter still in Fe~XIV, but there is no obvious signal in the 160~km/s asymmetry map there. There are regions of high velocity blue asymmetries (suggestive of upflows) present in the region prior to the eruption however and perhaps the most prevalent is that at $[660\arcsec, -130\arcsec]$ which is on the darker loops coming out of the North of the sunspot where we have seen similar multi-thermal outflow signatures before (cf., \opencite{2009ApJ...693.1306M}). We also note the clear signature of hot (Fe~XIV) upflows in the negative polarity region at the southern portion of the active region $[630\arcsec, -160\arcsec]$ while it is puzzling no clear signature exists in the Fe~XIII emission.

After the eruption, when the corona behind the CME is open, we see that the asymmetry maps have changed considerably, signatures of strong upflows are now seen in the flux concentrations at the bottom of the dimming region and over more of the active region. The most conspicuous of the new upflow regions in both spectral lines at 110~and 160~km/s is rooted in location 2, now at $[560\arcsec, -150\arcsec]$, the small flux concentration $50\arcsec$ to its East and the positive polarity region at the other end of the dimming region $[740\arcsec, -220\arcsec]$. 

In both lines, but certainly clearer in 274\AA{}\footnote{We suspect that, while the 202\AA{} Fe XIII emission line is spectrally clean according to the EIS spectral atlas (\opencite{2008ApJS..176..511B}), there may well be a subtle, undocumented blend or background issue in the red wing of the line that is affecting the R-B analysis. Regardless of this issue, there is dominant blue wing signature in the magnetic regions post-eruption that is greater in magnitude than this ``contamination". We are aware of the issue with this line and are exploring multiple EIS datasets to identify what is going on.}, the southern portion of the sunspot region (e.g., $[670\arcsec, -170\arcsec]$) is showing larger contiguous regions of upflow. Interestingly, this location was studied in detail by \inlinecite{2009arXiv0901.2817M} and it was noted that the penumbral structure of the region disappeared after the eruption, not to return while the region was visible on the disk. This observation will enter into our deliberations of what we are seeing and why (below). We also note that, in Fe~XIII, we still see the weak upflow signal from location~1 although it appears to become increasingly extended as the active region complex rotates towards the limb. 

Of course, in addition to the upflows that become visible over the course of the event, we see high speed hot downflows in certain locations. These are most clearly idenified as the bright red regions (e.g., at $[700\arcsec, 190\arcsec]$) most easily seen in Fe~XIV. We see that these downflows occur at the bottoms of the bright, compact, post-eruption loops that form as the corona starts to close behind the CME. Incidentally, these downflows are reduced in the later phase of the event.

This analysis, maps and associated movies, provide further interesting insight into the reaction of the coronal plasma to the morphology change imposed on it by the eruption of the CME. In the following section we will attempt to place this analysis in context and provide some insight into the possible impact of the on-disk changes to the CME itself.

\begin{figure} 
\centerline{\includegraphics[width=1.0\textwidth,clip=true]{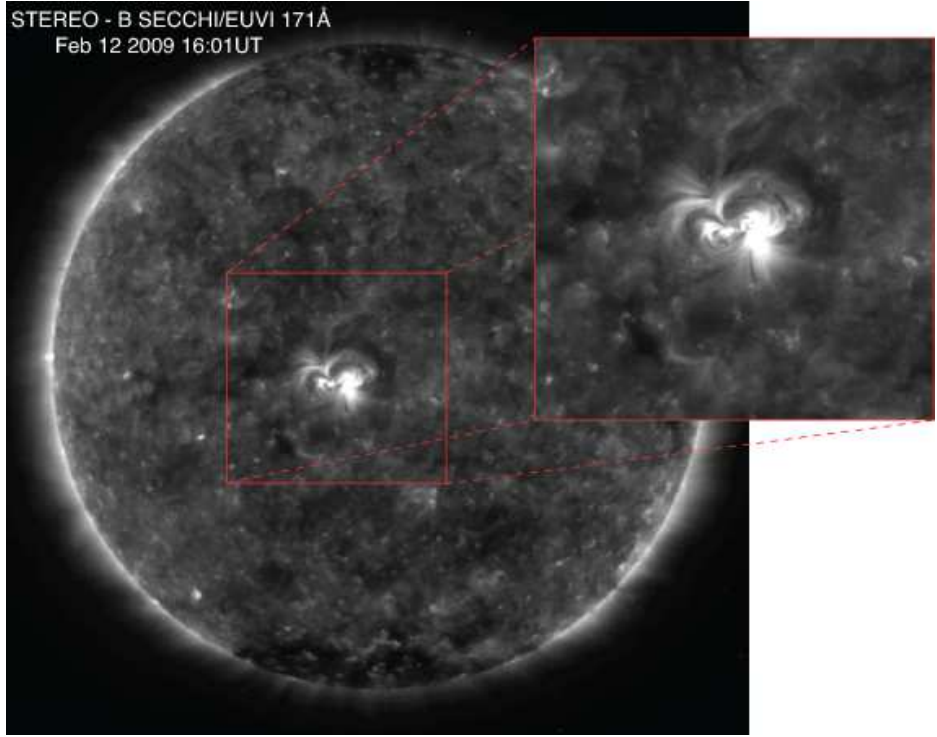}}
\caption{{\em STEREO} SECCHI/EUVI B image taken at 16:01UT on February 12, 2009. The movie that accompanies this figure shows four hours of evolution in the small active active region in the figure inset at a cadence of 95s. The signature of the plasma ``blobs'' is clearly visible in the movie as a bursty outflow that appears to travel along the magnetic loop structures that comprise the AR. These blobs are visibly traveling outward from the surface before and throughout the eruption.}\label{fig5}
\end{figure}

\subsection{STEREO Images the Relentless Outflow}
As has been demonstrated by \inlinecite{2009arXiv0910.2452M} the significant line profile asymmetries observed here have been connected with the visual appearance of plasma ``blobs'' that are visible in broadband coronal imaging diagnostics from {\em TRACE}, {\hinode}/XRT and the {\em STEREO\/} spacecraft. In Fig~\pref{fig5} and the accompanying movie (Movie~4) we see a small, asymmetric coronal dimming event that was observed in the Fe IX/X 171\AA{} passband by SECCHI/EUVI (\opencite{2008SSRv..136...67H}) on {\em STEREO}~``Behind'' on February 12, 2009 as part of a special quadrature observation sequence. During this sequence a 171\AA{} image was taken every 95s. The purpose of including this intriguing observation in the present paper is two-fold. It highlights that the energy release into the corona, solar wind through the magnetic regions of the lower atmosphere is relentless, and that EUV imagers can provide some valuable information about more than the morphology of the coronal plasma. 

The small active region near the center of the disk (NOAA AR~11012) was a prolific source of plasma blobs, which, as \inlinecite{2009arXiv0910.2452M} has demonstrated, are rooted in the same location as the multi-thermal coronal emission line asymmetries, and dynamic Type-II spicules that display similar velocity characteristics. It is fortunate that while spectroscopic instruments like EIS are burdened by the line-of-sight sensitivity of the weak plasma emission due to the spicule driven mass outflow, broadband imagers do not suffer the same effects, as spicule driven events are unlikely to be Doppler shifted out of the passband. These discrete, quasi-periodic mass loading events are visible on most coronal loops provided that the net brightness of the those loops is low enough (they can be long, or low density): then the 5\% quasi-periodic mass injections are quite easily visible on the loop system. 

Prior to the eruption of the AR (at 18:07UT) we readily see the relentless release of hot mass heading upward into the corona above. As the corona is opened by the small filament eruption, the loop system to the East of the small active region is forced outward. The locations tethering those stretched out field lines continue to show the transmission of blobs throughout the dimming event. 

We expect that it may be possible to use this form of EUV imaging data to study in some detail the energetics and (quasi-)periodicity of the mass loading events occurring before, during and after an event. Our suspicion is that the quasi-periodicity of the mass-loading events on those magnetic field lines and the apparent association of these events with Alfv\'en waves present an environment in which Alfv\'en wave dissipation can drive the resulting fast wind stream behind the CME. A similar assertion is made when considering blob-like activity in polar plumes and their impact on the fast wind in polar coronal holes (\opencite{2009McIntoshPloomz}). This is discussed in more detail below.

\section{Discussion}\label{s:discuss}
It is clear from the discussion above that the R-B maps of the coronal plasma change dramatically over the course of the event studied. We should stress that places where upflows are seen after the eruption are also likely to be sources of upflow {\em before} the event (an assertion validated by the {\em STEREO} observations, albeit of a different event) and this factor is critical to understanding what we observe. The reason for this apparent ``switch-on'' in the spectroscopic measurements must be largely geometric, the viewing angle between the magnetic field direction (along which the flows occur) and the line-of-sight is critical for determining the appearance of these weak blue Doppler asymmetries in relation to the bulk of the line emission. On highly inclined field lines to the observer's line-of-sight we will see little of the field directed motion, $\cos\theta$ is small and the true velocity of the upflow component is shifted towards lower velocities by that factor, with the end result that the upflow emission becomes part of the bulk of the line, rendering it practically invisible to our analysis\footnote{A study of this active region complex crossing the solar disk, and its impact on R-B analysis, will be completed shortly and be presented in the literature.}. Conversely, when the observer is looking straight down on one of these field lines, we are able to see both the peak (weak) emission and the blue wing contribution. Depending on the relative magnitude of the blue wing contribution, the net effect of fitting the line profile with a single Gaussian profile is a larger $1/e$ width (as discussed in McIntosh et al. 2009b) and an additional weighting of the line centroid to the blue by a few km/s. This additional profile broadening in the magnetic footpoints of the dimming region (in terms of a single Gaussian fit) augments the line broadening reported by \inlinecite{2009ApJ...693.1306M}. We speculate that we may have a two-stage process of mass-loading at the bottom of those field lines and release of Alfv\'en waves driven by the generation process of the mass-loading events and/or the change in tension along the field line due to the mass loading. These waves then propagate outward.

It is important to note that the emission observed by EIS over these strong magnetic field regions has {\em at least} two components. The R-B analysis we present is just one way of validating the weaker outflow component, and one that does not prescribe a particular functional form to the distribution observed. While the excess wing emission looks, to all intents and purposes, Gaussian in nature we believe that multi-Gaussian fits to the data are informative, validate the presence of the high-velocity component, but offer little in terms of the complete physical description of these potentially important velocity distributions.

\begin{figure} 
\centerline{\includegraphics[width=1.0\textwidth,clip=true]{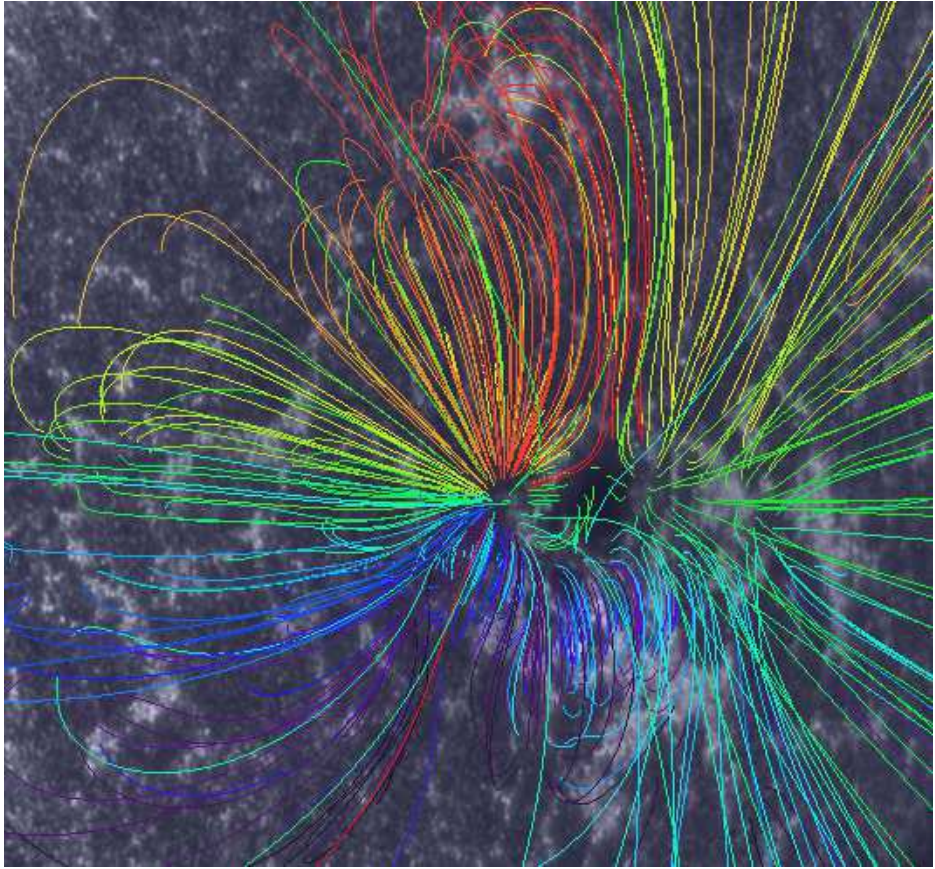}}
\caption{A snapshot of the potential field extrapolation of the active region studied before the eruption. We have plotted the magnetic field lines that are rooted in the blue dashed region of Fig.~\pref{fig1}. The background image is that of the {\em TRACE} 1600\AA{} passband and shows the mixture of dark spot structure and brighter plage emission distributed around this complex AR.}\label{fig4}
\end{figure}

A snapshot of the magnetic field geometry of this active region prior to the eruption is shown in Fig.~\pref{fig4}. Using the composite MDI/SP line of sight magnetogram shown in Fig.~1 we compute the associated potential field and plot some of the field lines rooted in the blue-dashed region to get a picture of magnetic connectivity in the region. Unfortunately, it is unlikely that this line-of-sight magnetogram evolution offers much insight into the subtleties of the field geometry change following the CME. While there is a clear impact on the chromospheric (Ca~IIH) and photospheric (G~band) emission in the vicinity of the sunspots (as reported in \opencite{2009arXiv0901.2817M}), and in their penumbral structure following the eruption (as we have noted above), the spectro-polarimetric signal relative to that change is subtle and, to all practical purposes, hard to decipher (especially given the proximity of this region to the limb). The complexity of Stokes profile interpretation, and the resulting field representation, in the SP measurements is a topic we will leave to a more detailed subsequent investigation. 

In light of the difficulties with interpreting the SP data we suggest to consider our R-B maps as an interpretative guide to the magnetic geometry. So, as the coronal magnetic field lines open up in response to the eruption of the CME we start to see a larger component of the upflow rooted ubiquitously in strong magnetic flux concentrations \cite{2009ApJ...701L...1D,2009arXiv0910.2452M} along the line-of sight observed with EIS. These outflows on the now open magnetic field lines are clearly oriented {\em behind\/} the CME, the plasma is hot ($>$1MK) and, as such has enough thermal pressure to overcome the Sun's gravitational field on those field lines (\opencite{1991ApJ...372..719P}) and is, to all intents and purposes, a stream of solar wind. The strength of the magnetic field at the bottom of these open regions is considerably larger than that of typical (equatorial) coronal holes (\opencite{2006ApJ...644L..87M}) where ubiquitous outflows are readily observed in the magnetic network (\opencite{2009ApJ...701L...1D}; McIntosh, Leamon \& De Pontieu, in preparation) and may go some way to explaining why some coronal dimming events rooted in very strong magnetic flux regions have {\em very\/} fast CMEs that show little or no deceleration and have associated high wind streams behind them (e.g., \opencite{1997JGR...10219743N}; \opencite{2004JGRA..10909102S}).

Based on the analyses of \inlinecite{2009ApJ...701L...1D}, \inlinecite{2007Sci...318.1574D} and \inlinecite{2009ApJ...693.1306M} we suggest that the spicules that form the root of these upflow regions transport mass and Alfv\'{e}nic wave energy behind the CME. It is expected that the significant Alfv\'{e}nic energy present on the spicules, and coronal magnetic field lines, will propel the plasma outward from the Sun (e.g., \opencite{2006JGRA..11106101S}; \opencite{2007ApJS..171..520C}; \opencite{2007ApJ...662..669V}). The presence of the mass and energy flux behind the CME poses an interesting challenge for our current understanding of CMEs, and particularly those that do not slow down from the $\sim$1000~km/s initial plane-of-the-sky observed speeds to the ambient wind speed of a few hundred kilometers per second through interplanetary space (e.g., \opencite{2009ApJ...705..914R}): Does the quasi-periodically forced stream of mass and energy in the observed upflows have a significant impact on the momentum balance of the CME such that it would provide a continuous ``push'' for the material ahead to overcome the inertia of the plasma that the CME is propagating through?

In keeping with the tone of this article as one that is looking to future diagnostics of surface activity and their net impact on space weather, we estimate that the Atmospheric Imaging Array (AIA) of the {\em Solar Dynamics Observatory} (SDO) will have increased signal-to-noise over SECCHI/EUVI of a factor of several. Coupled with the high image cadence (10s), multi-thermal observations, and spatial pixels that are a factor of three smaller ($\sim$0.6\arcsec), we are entering a phase of solar physics when we will be able to investigate the root energetics of the outer solar atmosphere and the possible impact those have on associated ejecta. We anticipate that these observations can, and will, provide ideal boundary information for the increasingly sophisticated space weather modeling effort. In late 2012, IRIS will provide unprecidented spectra of these events in the transition region and upper chromosphere that will complement the observations of SDO/AIA and {\em Hinode} (if it is still operational) in such a way that we will be able to provide even stronger observational constraints towards resolving the puzzle of CME acceleration.

\section{Conclusion}\label{s:summary}
We have observed the ``triggering'' of high-velocity coronal outflows behind a CME that are rooted in the field of newly open strong magnetic flux regions. These upflows carry mass and Alfv\'{e}nic wave energy outward from the Sun and, as such, are a potential momentum source for the CME while the coronal magnetic field is open. 

We have observed a rapid switch of changes in ``surface'' thermodynamics from plasma heating to plasma forcing on the open magnetic topology and the relentless nature of the mass loading on those magnetic field lines at all phases of the eruption. Clearly, a larger sample of spectroscopically studied CMEs (preferably from high-cadence, full-disk spectral imaging instrumentation of the chromosphere and corona) are needed to study the true impact of this potential driver. While this paper is very speculative in nature, and we offer no solution to the issue at this point (sorry), the observations presented, and results discussed, give significant food for thought. In the near future observations such as those discussed herein will help us ascertain if the resulting fast solar wind outflow behind the CME impacts the kinematics of the disturbance as it travels into the inner heliosphere.

%
\begin{acks}
SWM is supported by NSF ATM-0541567, NASA NNG06GC89G; BDP by NASA grants NAS5-38099 (\trace), NNM07AA01C ({\em Hinode}) and NNG06GG79G. SWM and BDP are jointly supported by NASA grants NNX08AL22G and NNX08AH45G. RJL is partially supported by NASA grant NNH08CC02C.
\end{acks}

%
%
\bibliographystyle{spr-mp-sola}
\bibliography{mcintosh}  


\end{article} 
\end{document}